\documentclass[twocolumn,amsmath,amssymb,aps,prb]{revtex4-1}

\usepackage{graphicx}
\usepackage{dcolumn}
\usepackage{bm}
\usepackage{array}

\usepackage[utf8]{inputenc}
\usepackage[T1]{fontenc}
\usepackage{subcaption}

\begin{document}

\title{Mott and spin-Peierls physics in TiPO${}_4$ under high pressure\\}

\author{H. Johan M. Jönsson\textsuperscript{1}}
\author{Marcus Ekholm\textsuperscript{1,2}}
\author{Silke Biermann\textsuperscript{3,4,5,6}}
\author{Maxim Bykov\textsuperscript{7,8}}
\author{Igor A. Abrikosov\textsuperscript{1,9}}

\affiliation{{}\textsuperscript{1}Theory and Modelling, IFM-Material Physics,
Linköping University, SE-581 83, Linköping, Sweden}
\affiliation{{}\textsuperscript{2}Swedish e-Science Research Centre (SeRC),
Linköping University, SE-581 83, Linköping, Sweden}
\affiliation{{}\textsuperscript{3}CPHT, CNRS, Ecole Polytechnique, Institut
  Polytechnique de Paris, F-91128 Palaiseau, France}
\affiliation{{}\textsuperscript{4}Coll\`ege de France, 11 place Marcelin Berthelot, 75005 Paris, France}
\affiliation{{}\textsuperscript{5}Department of Physics, Division of Mathematical Physics, Lund University, Professorsgatan 1, 22363 Lund, Sweden}
\affiliation{{}\textsuperscript{6}European Theoretical Spectroscopy Facility, 91128 Palaiseau, France, Europe}
\affiliation{{}\textsuperscript{7}Laboratory of Crystallography, University of
Bayreuth, 95440 Bayreuth, Germany}
\affiliation{{}\textsuperscript{8}Bayerisches Geoinstitut, University of Bayreuth,
95440 Bayreuth, Germany}
\affiliation{{}\textsuperscript{9}Materials Modeling and Development Laboratory,
 National University of Science and Technology ‘MISIS’, Moscow, 119049, Russia}

\begin{abstract}
    TiPO$_4$ is a Mott insulator and one of few inorganic compounds featuring a
    spin-Peierls phase at low temperature. Recent experimental studies have suggested the presence of
    spin-Peierls dimerization also at ambient temperature though at high pressure.
    Here, we present a combined experimental and theoretical study of the energetics of the high-pressure
    phase. 
    We analyse dimerization properties and their coupling
    to spin degrees of freedom.
    Most importantly, we argue that TiPO$_4$ presents a direct analogue
    to the celebrated binary transition metal oxide VO$_2$. TiPO$_4$
    allows to assess spin-dimer physics in the high-pressure regime in
    a controlled fashion, having the potential to become an important
    model system representative of the class of dimerized transition
    metal oxides.
\end{abstract}

\maketitle

\section{\label{sec:Intro}Introduction}\noindent
Compounds with strong electronic Coulomb correlations
are a hot topic of modern solid state physics.
Indeed, electronic correlations may lead to promising phenomena, such as
metal-insulator transitions, colossal magnetoresistance \cite{jrn:dagotto}
or high temperature superconductivity \cite{jrn:stepanov}.
Moreover, a theoretical description of the underlying interacting
many-body system is highly challenging.
Incomplete screening of electronic interactions can lead to short-lived
quasiparticles and breakdown of Fermi liquid theory.
A typical example is the Mott metal-to-insulator transition at low
temperature, in which the electron-electron repulsion energy dominates
over the kinetic energy, resulting in localisation of the electrons to the
atomic sites, and loss of conductivity.

Furthermore, the charge, spin and orbital degrees of freedom may be coupled to the lattice in a non-trivial
fashion. This may lead to highly interesting effects, especially in low dimensional materials,
such as the spin-Peierls transition. This effect may be observed in systems with
(quasi-) one-dimensional spin chains. CuGeO$_3$ was the first inorganic compound
showing this transition, which takes place below 14 K \cite{jrn:hase93}.
More recently, the spin-Peierls effect was observed in the Mott insulator TiOCl
at low temperature \cite{jrn:seidel03,jrn:shaz05}. Theoretical and experimental studies of TiOCl
reveal the strongly corelated nature of this compound and the need to go beyond the static mean-field
description of d-d interactions\cite{jrn:sahadasgupta05} as well as to include
non-local interactions to describe its electronic structure
\cite{jrn:sahadasgupta07,jrn:aichhorn09}. Remarkably the spin-Peierls effect in
TiOCl has been reported at temperatures up to 215 K when subjected to pressure\cite{jrn:rotundu18}.
These results were interpreted as an enhancement of the magnetic interaction
energy at high pressure due to increased overlap between the adjacent d orbitals.

\begin{figure}[ht]
    \centering
    \begin{subfigure}{\linewidth}
        \includegraphics[trim={3cm 4cm 40cm 11cm},width=0.9\linewidth,clip]{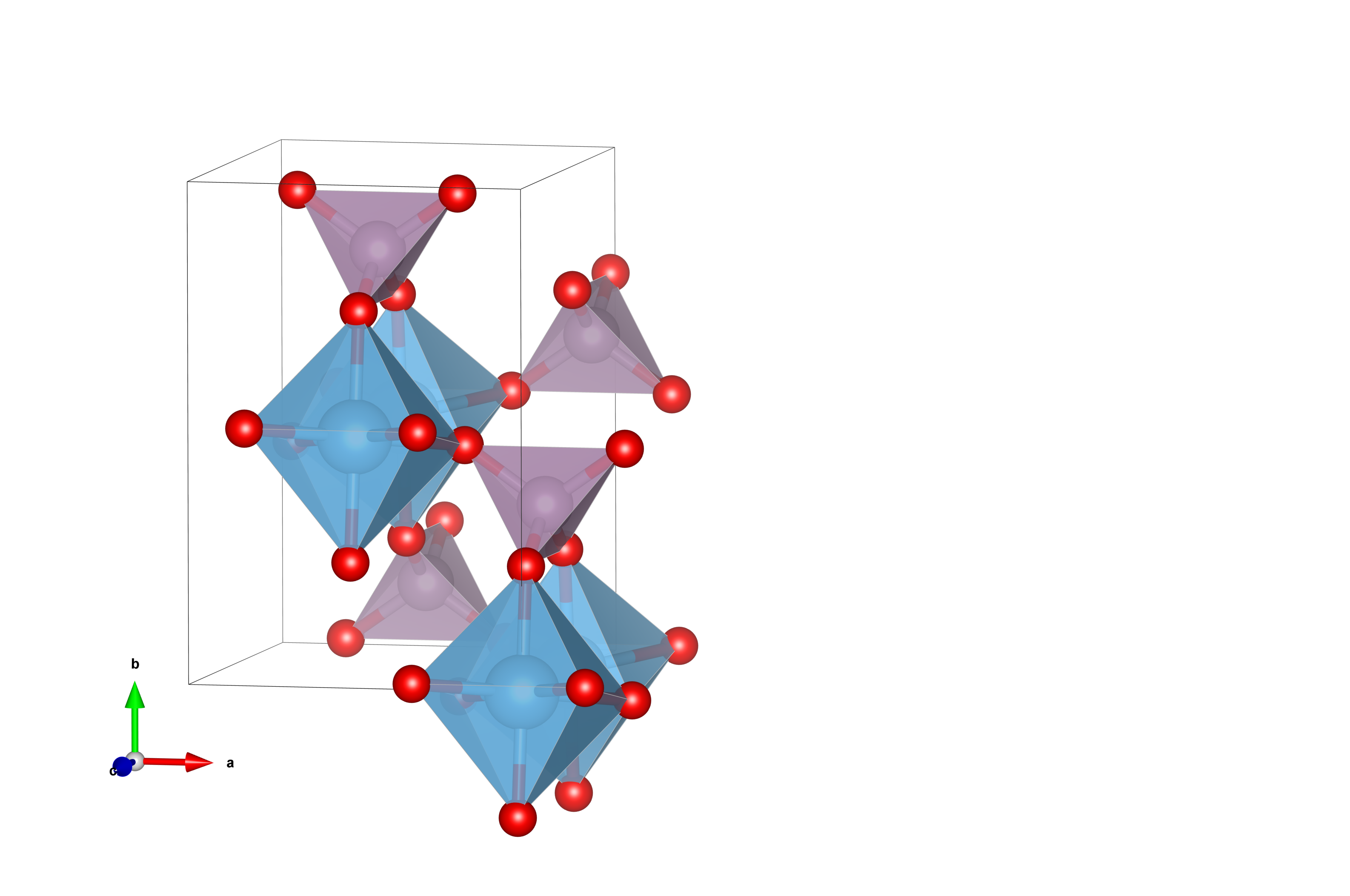}
        \caption{}
        \label{fig:cryststruct-I}
    \end{subfigure}
    \begin{subfigure}{\linewidth}
        \includegraphics[trim={7cm 5cm 35cm 14cm},width=0.9\linewidth,clip]{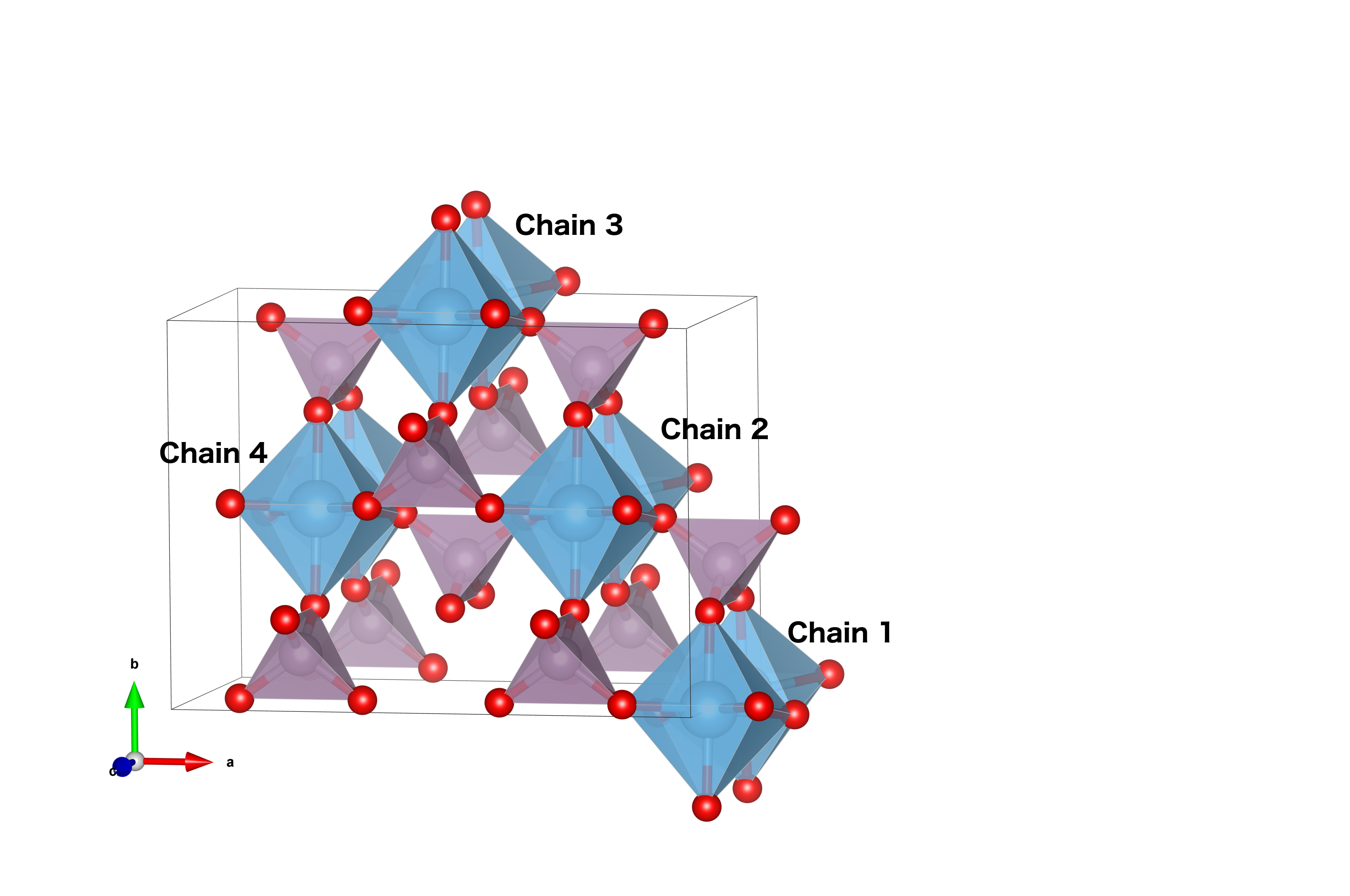}
        \caption{}
        \label{fig:cryststruct-III}
    \end{subfigure}
    \begin{subfigure}{\linewidth}
        \includegraphics[trim={15cm 6cm 40cm 6cm},width=0.5\linewidth,clip]{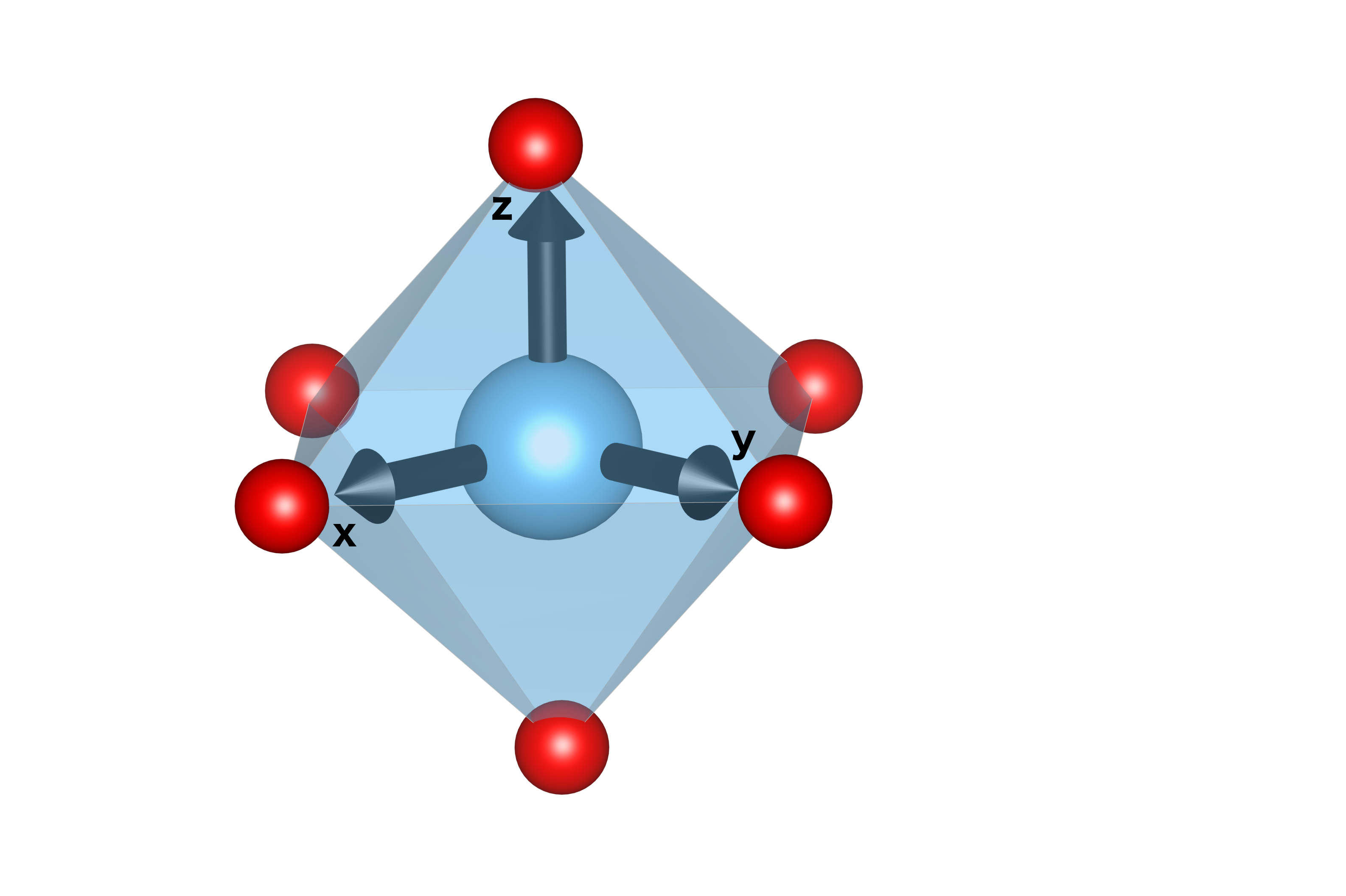}
        \caption{}
        \label{fig:localbasis}
    \end{subfigure}
    \caption{Crystal structure of a) phase I (space
    group \textit{Cmcm}) and b) phase III (space group \textit{P2$_1$nm}) of TiPO${}_4$.
    Ti atoms are shown in blue, P are purple and O are red. Also shown in blue
    are the Ti-O complexes and the P-O complexes are purple. Panel c) shows a
     schematic picture of the local coordinate system (defined as $x,y, z$)
     around a given Ti atom used for the calculation of projected density of
     states. The Ti chains lie along the $x+y$ line.}

    \label{fig:cryststruct}
\end{figure}

In this work we study the TiPO$_4$ system, which also has been reported to undergo a
spin-Peierls transition, at room temperature, at high pressure \cite{jrn:bykov16}.
We investigate the interplay between the Ti dimerization and electronic
structure from ab-initio calculations as well as experiments.

At ambient conditions, TiPO$_4$ crystallizes in the CrVO$_4$ structure (space
group \textit{Cmcm}), referred to as phase I.
Tilted, edge-sharing TiO$_6$ octahedra form quasi-one-dimensional chains along the $c$-axis,
interconnected by PO$_4$ tetrahedra (Fig. 1a).
Early measurements of magnetic susceptibility\cite{jrn:kinomura82} and neutron
diffraction scattering\cite{jrn:glaum96} failed to detect magnetic ordering down to 2 K.
Subsequent susceptibility measurements combined with nuclear magnetic resonance
(NMR)\cite{jrn:law11} indicated
antiferromagnetic
coupling along the chains, with a singlet formation below 74 K:
the Ti atoms form dimers within the chains, with alternating Ti-Ti
distances, $d\pm\delta$, a phenomenon interpreted as a spin-Peierls transition
\cite{jrn:bykov13}.

When subject to pressure above 4.5 GPa at room temperature, TiPO$_4$ enters
the incommensurately modulated crystallographic phase II.
At 7 GPa, the commensurate phase III (space group \textit{P2$_1$nm}) sets in,
which is a foutfold superstructure
of phase I, where the Ti-atoms again form dimers.
The TiO$_6$ octahedra are seen to be tilted with respect to the crystallographic b axis.
From X-ray diffraction experiments it has been found that among the octahedra chains, there
are 4 different tilt angles, ranging from $5^\circ$ to $9^\circ$ (for
experimental details see supplementary information).
In this sense, one may speak of 4 different Ti chains.
As can be seen in Fig. \ref{fig:cryststruct}
 chains 1 and 3 lie in
the same [010]-plane and are equally dimerized. However, the Ti-Ti distances
$d \pm \delta$ are modulated out of phase with respect to each other. This is
also true for chains 2 and 4. Reminiscent of phase I, the observed dimerisation
has therefore been proposed as an extraordinary case of a spin-Peierls transition
occuring at room temperature \cite{jrn:bykov16}.

In fact, the physics of TiPO$_4$ is in many respects reminiscent of that of VO$_2$,
a time-honored and particularly intriguing example of a transition metal
oxide. The binary oxide VO$_2$ displays a first-order metal-insulator
transition as a function of temperature, switching from a high-temperature
bad metal phase to a low-temperature insulator.
While VO$_2$ was early on invoked as an example of a Mott transition
\cite{jrn:mottreview68}
, it became clear over the years that the accompanying
structural changes -- from a high-temperature rutile to a low-temperature
monoclinic phase -- also played a role in the transition, opening a
decade-long debate \cite{jrn:goodenough71,jrn:eguchi08,jrn:eyert02}.
Indeed, the structural transition proceeds via a doubling of the unit
cell along the crystallographic c-axis and a dimerization and slight
tilting of vanadium atoms along chains in that direction.
Cluster Dynamical Mean Field Theory calculations
\cite{jrn:biermann05,jrn:tomczak08} characterized the phenomenon as
"correlation-assisted
Peierls transition", establishing that the insulating phase emerges
from the metallic one thanks to (1) a rearrangement of charge within
the t$_{2g}$ manifold that leaves the single electron per V atom fully in
the a$_{1g}$ orbitals and (2) the formation of a spin singlet ground
state in the bonding combination of these a$_{1g}$ states of neighboring
vanadium atoms in the dimers. The resulting picture is in agreement
with the measured flat magnetic susceptibilities in the insulating
phase \cite{jrn:pouget76} and is confirmed by photoemission spectroscopy
\cite{jrn:koethe06}.

Investigation of VO$_2$ remains  an active field of research for
several reasons: (1) The transition takes place at temperatures of
around 340 K, that is, slightly above room temperature, making the
compound an ideal material for applications exploiting the metal-insulator
transition. In this respect, it is particularly interesting that the
transition temperature can be systematically lowered by tungsten
substitution \cite{jrn:tan12}, although the precise mechanism of this effect
is not yet understood. (2) Despite the undoubtable role of the structural
distortions in the transition, various variations of the phenomenon
indicate the simultaneous presence of strong electronic correlations.
Pouget et al. pointed out \cite{jrn:pouget76}, that uniaxial
stress or tiny amounts of Cr-doping modify the distortions -- in
particular leaving half of the V atoms undimerized -- while still
inducing an insulating phase. In the same vein, pump-probe experiments
suggest that structural and electronic transitions can be
decoupled \cite{jrn:jager17}. Finally, introducing oxygen vacancies into the system
eventually suppresses the transition altogether \cite{jrn:fan18}.
All these elements suggest VO$_2$-derived materials or -- more generally
-- materials where dimerization of transition metal ions plays a role
are still hiding rich and interesting -- and potentially useful --
physics. It is therefore intriguing to have, with TiPO$_4$, another
material at hand sharing various aspects of dimerization physics with
VO$_2$.


\section{\label{sec:Method}Methodology}\noindent
Calculations were performed using density functional
theory \cite{jrn:hohenbergkohn, jrn:kohnsham}, with the
projector augmented wave (PAW) method \cite{jrn:blochl94} implementation in the Vienna ab
initio simulation package \cite{jrn:VASP1,jrn:VASP2} (VASP). For the
exchange-correlation functional the local density approximation, with
on-site interaction (LDA+U) as parametrized by Dudarev et al. \cite{jrn:lda+u} was used,
we chose U = 2.0 eV, as will be motivated in section \ref{sec:Res}.

In order to achieve good convergence with regards to the plane wave energy an
energy cut-off of 520 eV was chosen. For the Brillouin zone integration we employed
a Monkhorst-Pack k-point grid of dimension $5\times7\times11$, resulting in a
total of 112 k-points in the irreducible Brillouin zone. The unit cells used for the
calculations, shown in Fig. \ref{fig:cryststruct}, contained
4 and 8 formulae units
respectively.

Details on the experiments are provided in the supplemental materials\cite{suppmat}.

\section{\label{sec:Res}Results and discussion}\noindent
\begin{table}
    \caption{Comparison of calculated equilibrium, equilibrium volume V$_0$, bulk modulus B$_0$ and its pressure derivative B', parameters and their room temperature
     experimental values}
    \begin{ruledtabular} \begin{tabular}{cccc}
	    & V${}_0$ [Å\textsuperscript{3}/atom]  & B${}_0$ [GPa] &
        B'\\
        \hline
    	LDA	& 10.37	& 109.92	& 4.97\\
	    LDA+U, U = 2.00 eV	& 10.78	& 87.12	&  5.39\\
        LDA+U, U = 3.00 eV	& 10.94	& 89.26	&  4.90\\
	    Experiment \cite{jrn:bykov16}& 11.16 & 72	& 6.5\\
    \end{tabular}\end{ruledtabular}
    \label{tab:Eq-param}
\end{table}\noindent

\begin{figure}[ht]
    \centering
    \includegraphics[width=\linewidth]{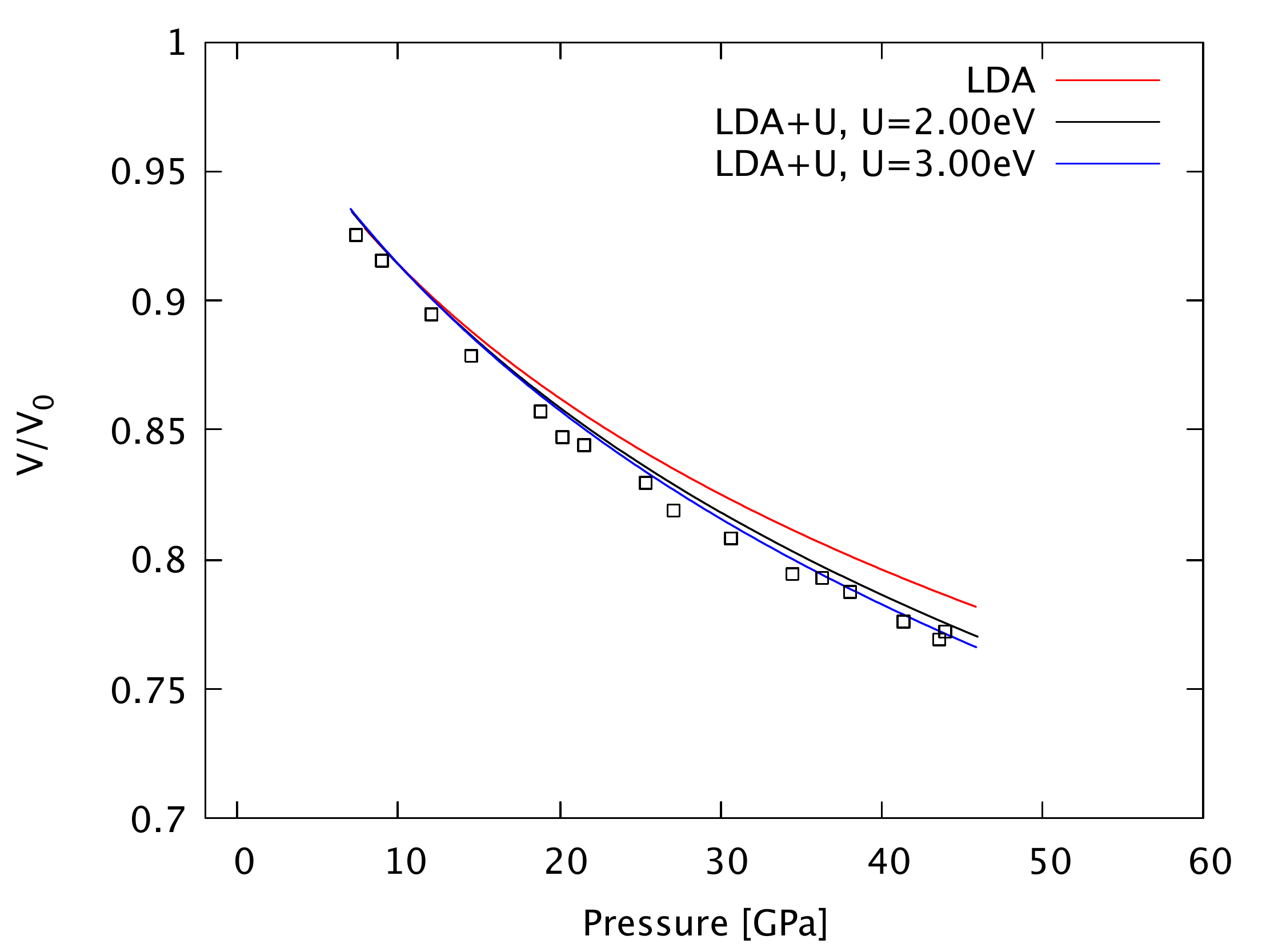}
    \caption{Equations of state calculated using LDA, LDA+U with
    U$=2.00$ eV and LDA+U with U=$3.00$ eV. Experimental results \cite{jrn:bykov16} are
    marked with open squares. Volume per atom, V, is shown relative to the equilibrium volume V$_0$. Calculations use the calculated zero pressure volume for $V_0$ and experiments the experimentally obtained zero pressure volume.}
    \label{fig:eos}
\end{figure}
\begin{figure}[ht]
    \centering
    \begin{subfigure}{\linewidth}
    \includegraphics[width=\linewidth]{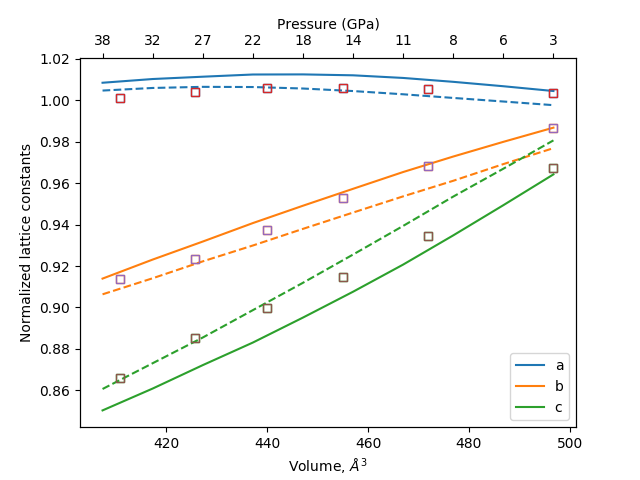}
    \end{subfigure}
    \caption{Calculated pressure dependence of the lattice constant  a, b, and c normalized by their calculated values at zero pressure. Solid lines show values obtained using LDA + U with U = 2.0 eV, dashed lines show values obtained with U = 3.0 eV. Open
     symbols are experimental values \cite{jrn:bykov16}.}
    \label{fig:normLat}
\end{figure}

\noindent
First, we have calculated the total
energy as a function of unit cell volume, relaxing the ionic positions at each volume.
The equilibrium
parameters obtained from our fit to the Birch-Murnaghan equations of state are
summarised in Tab. \ref{tab:Eq-param}, also plotted in Fig. \ref{fig:eos}. As
can be seen LDA underestimates the equilibrium
volume of the unit cell and slightly
overestimates the bulk modulus. Using LDA+U we improve V$_0$ and B$_0$.
The pressure dependence of the lattice parameters is plotted in Fig.
\ref{fig:normLat}.
We find that LDA + U with U = 2.00 eV shows best agreement with
experiments.

We note that according to our calculations at the theoretical equilibrium volume chains 1 through 4 in
figure \ref{fig:cryststruct-III} become eqiuvalent, and the optimized  structure
therefore becomes equivalent to phase I. At this volume
the Ti - O - Ti angle is 94.5$^{\circ}$. For comparison, we have carried out calculations at the experimental room temperature volume (see Table \ref{tab:Eq-param}) and
have obtained the value 96.5$^{\circ}$. This can be compared to the experimental
 value 95.5$^{\circ}$.

At the experimental volume the calculated
magnetic moment 0.80$\mu_B$ can be compared to the value
0.71$\mu_B$ found in reference \cite{jrn:lopezmoreno} using PBE.
In figure \ref{fig:tot-dos} the total density of states (DOS) for different
pressures is shown. Using LDA+U we correctly reproduce an insulating state\cite{jrn:jönsson19}. We see
an insulating band gap of $\approx1$ eV formed among the d-states.
Note that calculations carried out within
LDA only yields a pseudo-gap opened due to AFM order.
The states are divided into  a low-binding energy Ti-d part and a high-binding
energy part, dominated by O-p states.
The Ti peak below the Fermi energy contains 1 electron per formula unit, which
means that the system is in a $d^1$ configuration.
 Figure \ref{fig:Ti-dos} shows the local Ti-3d density
of states
projected onto irreducible orbitals, Fig. \ref{fig:localbasis}. Degeneracy of the
e$_{g}$ and t$_{2g}$ states is
lifted due to the octahedral symmetry. Moreover, we see that the singly occupied
d$_{xy}$-level is pushed  below the d$_{yz}$ and d$_{xz}$-levels
as the octahedra are distorted. Figure \ref{fig:d-parchg}
shows the charge density of the d$_{xy}$ orbital.
At ambient pressure, we find that the chains do not dimerize. This observation
does not agree with experimental observations for phase I. The disagreement can
be explained by the fact that the increase of elastic energy associated with
dimerization is not compensated by the energy gain associated with formation of
spin singlets, because the latter is not accurately described with the present
exchange-correlation functionals.
Here we would like to point out that a similar observation has been reported  in
 Ref. \cite{jrn:pisani07} for TiOCl. At the same time, the overall pressure
 dependences of the atomic volume (Fig. \ref{fig:eos}) and lattice constants
 (Fig. \ref{fig:normLat})
 are correctly captured by our calculations. We therefore proceed with the
 analysis of the evolution of the electronic and magnetic properties of TiPO$_4$
 upon compression.

\begin{figure}[h]
    \includegraphics[width=\linewidth]{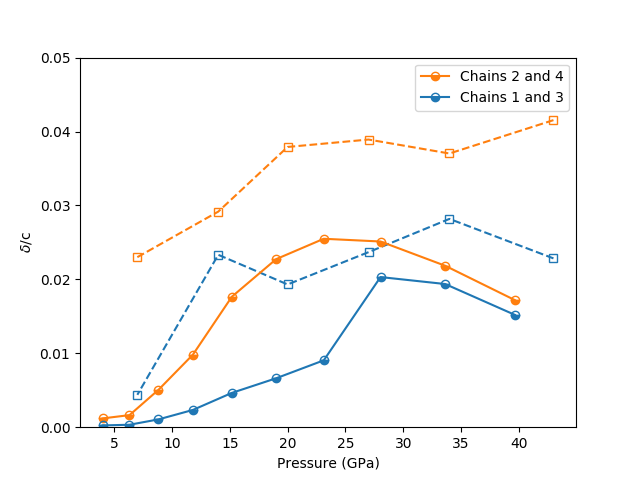}
    \caption{Dimerization along the c-axis, $\delta = d_{Ti-Ti} - c/2$, of
    chains 1 to 4 normalized by the c-lattice parameter. See Fig.
    \ref{fig:cryststruct} for the details of the crystal structure and
    designation of the chains. Experimental values are shown as open squares
    connected by dashed lines, calculated values are denoted by filled circles,
    for experimental details please see supplementary information\cite{suppmat}.}
    \label{fig:dim-AFM}
\end{figure}

\begin{figure}[ht]
    \centering
    \includegraphics[width=\textwidth]{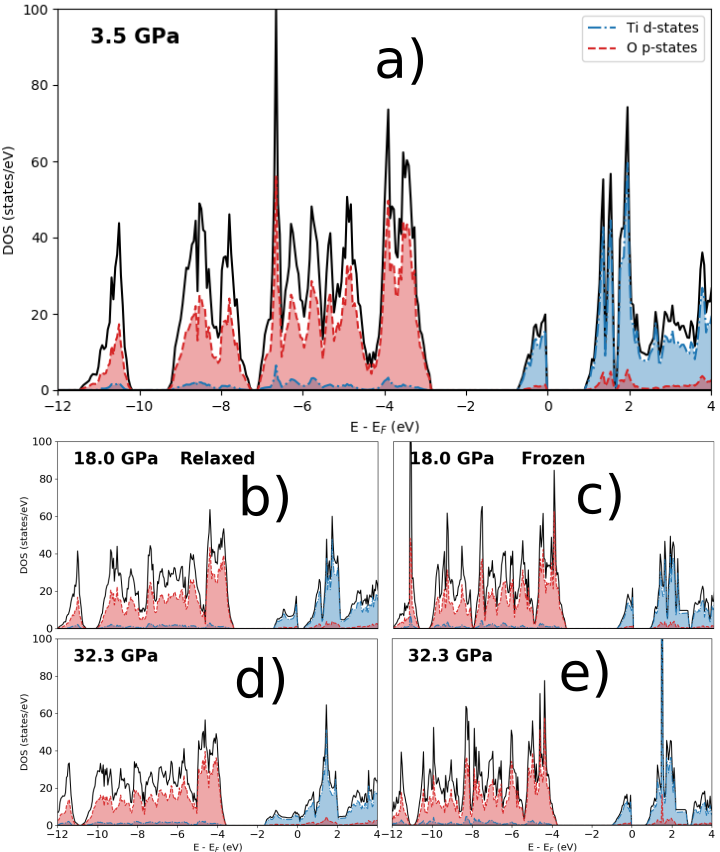}
    \caption{Total electronic density of states as a function of energy E, relative to the Fermi energy E$_F$, calculated at three different pressures, (a) 3.5 GPa, (b, c) 18.0 GPa and (d, e) 32.3 GPa. Panels (b) and (d) show results of fully relaxed calculations. Panels (c) and (e) show the results of calculations with ionic positions fixed to those obtained for the relaxed structure at 3.5 GPa.}
    \label{fig:tot-dos}
\end{figure}

\begin{figure}[ht]
    \centering
    \begin{subfigure}{\linewidth}
    \includegraphics[width=\linewidth]{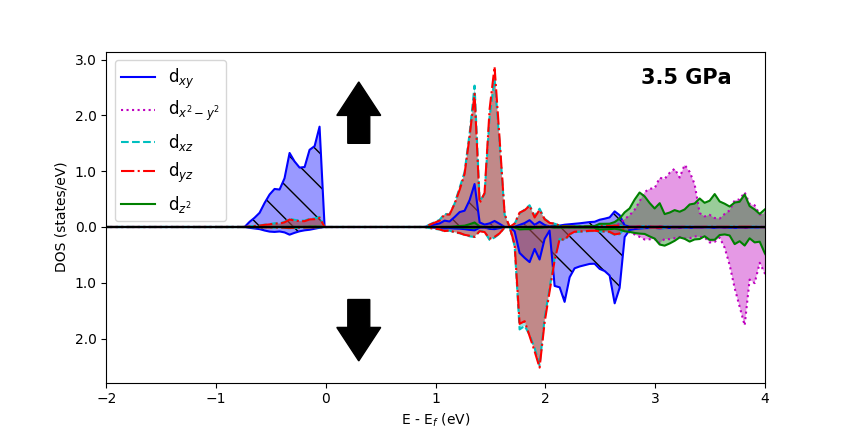}
    \caption{}
    \end{subfigure}
    \begin{subfigure}{\linewidth}
    \includegraphics[width=\linewidth]{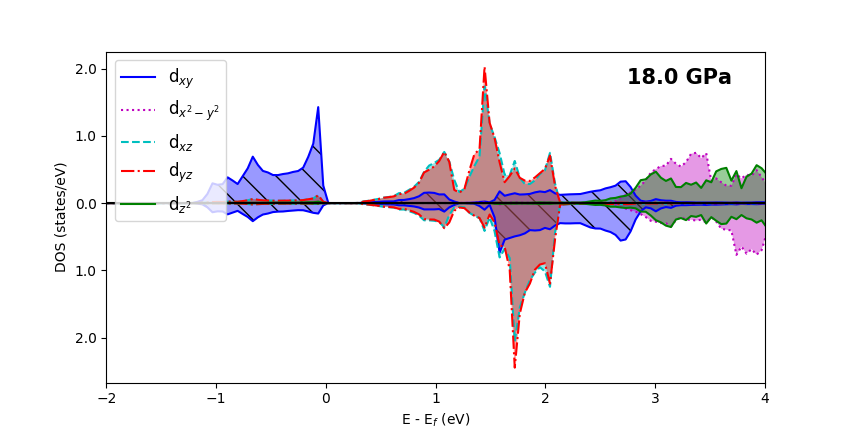}
    \caption{}
    \end{subfigure}
    \begin{subfigure}{\linewidth}
    \includegraphics[width=\linewidth]{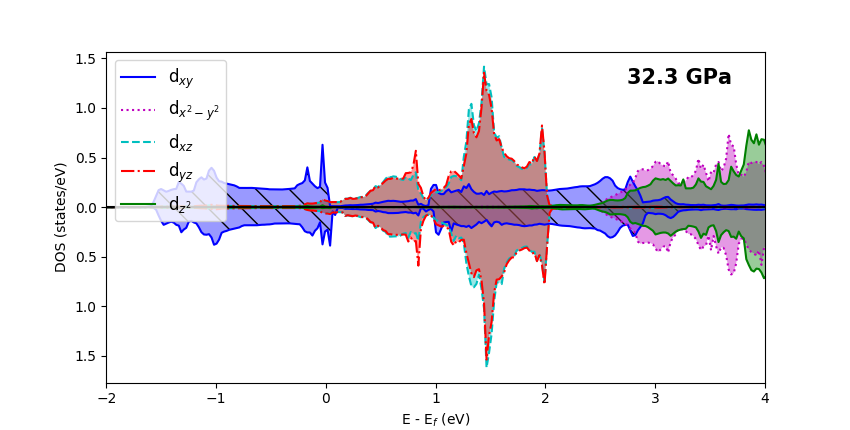}
    \caption{}
    \end{subfigure}
    \caption{Spin polarized electronic density of states projected onto the d-orbitals centered at a given Ti atom in the unit cell, as a function of energy E relative to the Fermi energy E$_F$, calculated at three different pressures, (a) 3.5 GPa, (b) 18.0 GPa and (c) 32.3 GPa. The spin up channel is depicted on the upper half of the y-axis and the spin down channel on the lower half.}
    \label{fig:Ti-dos}
\end{figure}

\begin{figure}
    \centering
    \begin{subfigure}{\linewidth}
        \includegraphics[width=\linewidth,trim={0cm 2cm 14cm 2cm},clip]{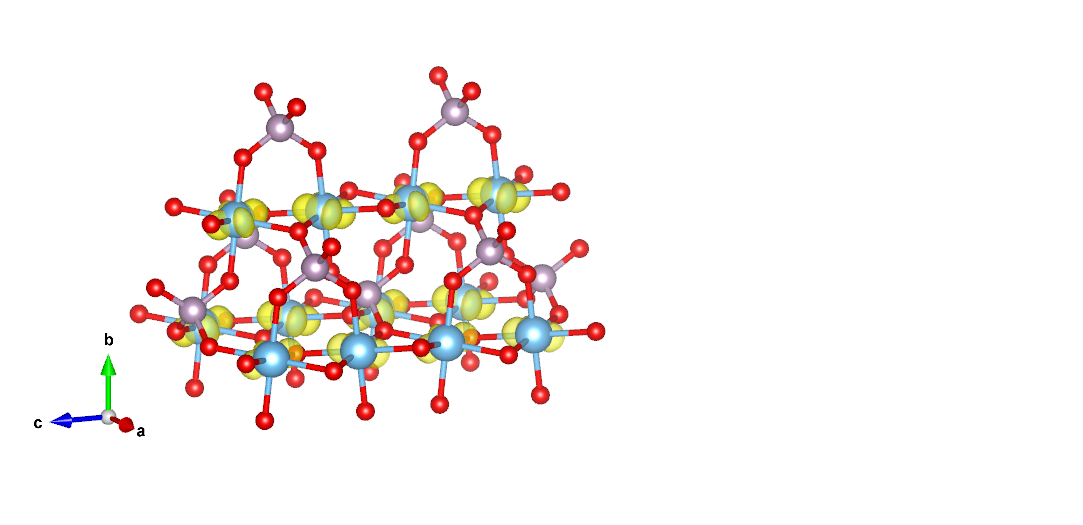}
        \caption{}
    \end{subfigure}
    \begin{subfigure}{\linewidth}
        \includegraphics[width=\linewidth,trim={0cm 2cm 14cm 2cm},clip]{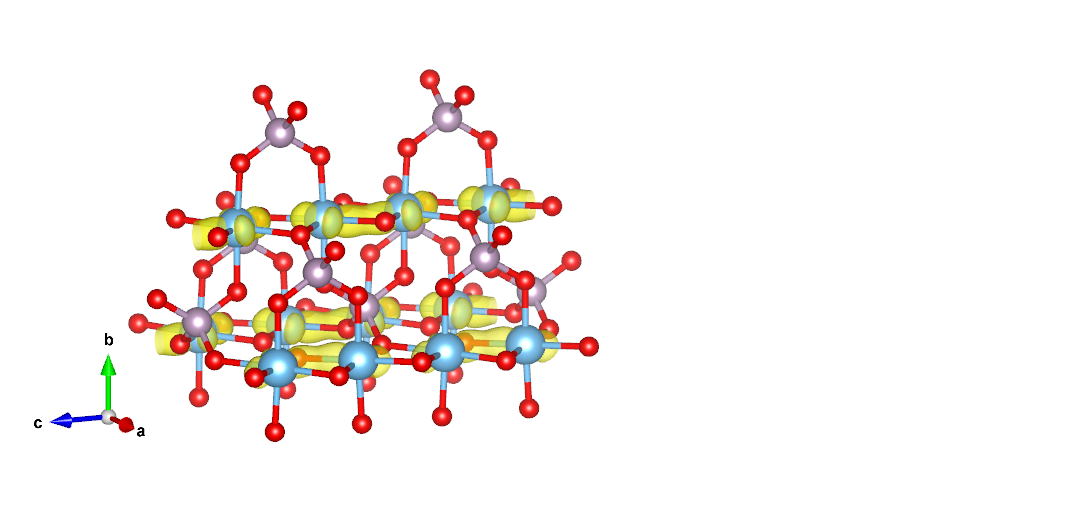}
        \caption{}
    \end{subfigure}
    \caption{Partial charge density of the d$_{xy}$-orbital calculated at pressures of (a) 3.5 and (b) 32 GPa.}
    \label{fig:d-parchg}
\end{figure}

\begin{figure}
    \includegraphics[width=\linewidth]{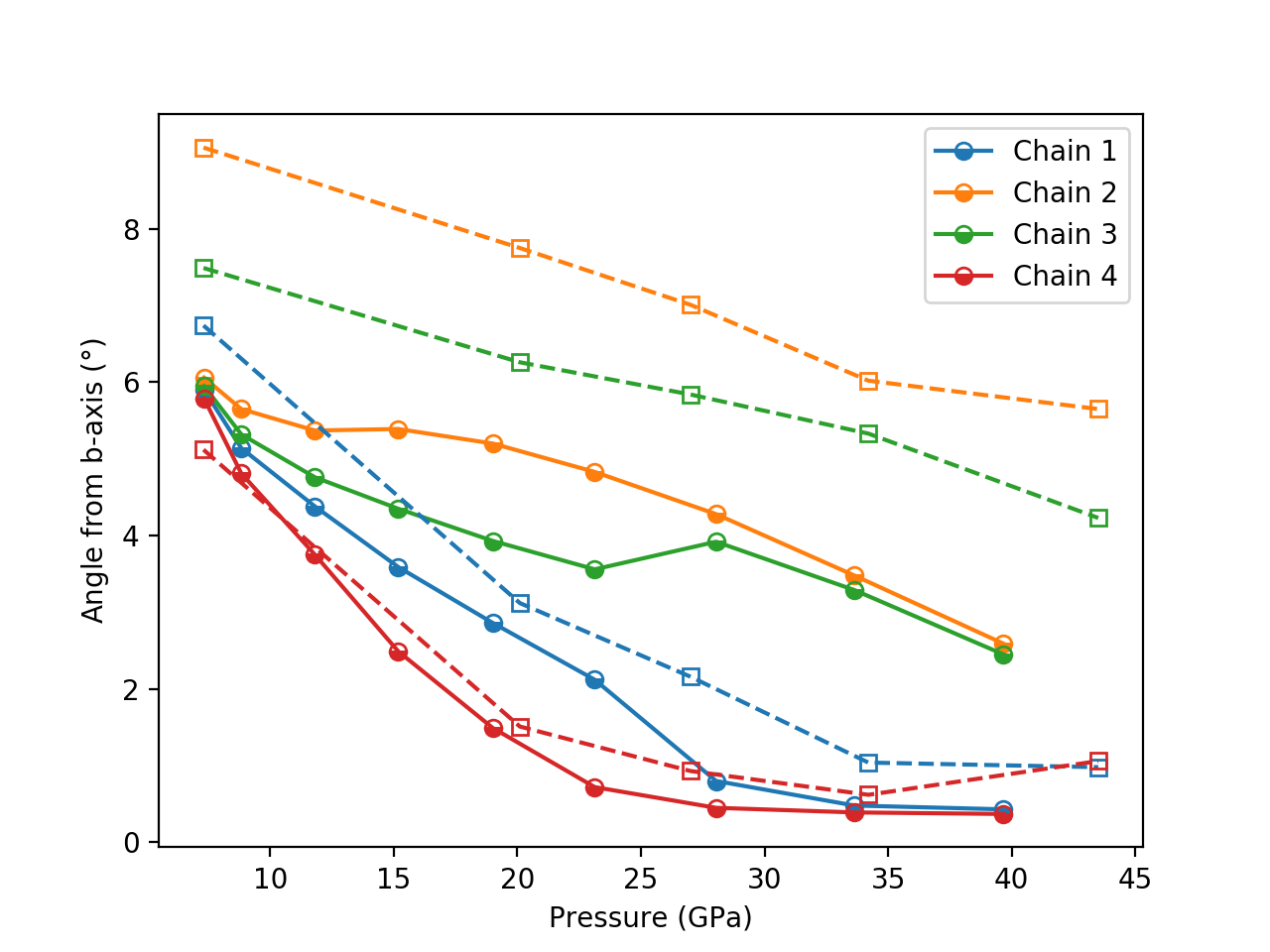}
    \caption{Angle between b-axis and octahedral major axis as a function of
    pressure for the four different Ti-chains. See Fig.
    \ref{fig:cryststruct} for the details of the crystal structure and
    designation of the chains. Experimental values (see supplementary
    information\cite{suppmat}) are shown
    with open squares connected by dashed lines, half filled circles denote
    calculated values.}
    \label{fig:angles}
\end{figure}

In Fig. \ref{fig:dim-AFM} we show the measured Ti dimerization as a function of pressure, along with computational results.
At the pressure where
phase III is observed experimentally we find that the chains indeed dimerize, although with a
somewhat reduced magnitude, as can be seen in Fig. \ref{fig:dim-AFM}.
At  the volume 0.94 V$_0$, which corresponds to a pressure of 6 GPa, the
four chains start developing distinct
tilting angles and thus become inequivalent, which is characteristic of the phase III
structure. The effect can be seen in Fig. \ref{fig:angles}. In our calculations the
transition from phase I into phase III occurs at around 9 GPa, slightly higher than
the experimentally observed transition pressure of 7 GPa.

As pressure is increased further we find that the b and c lattice parameters
decrease while the lattice constant a remains nearly constant (Fig. \ref{fig:normLat}). This difference in
compressibility along different crystallographic directions is in good agreement with
experiment \cite{jrn:bykov16}. Up to 23 GPa two trends can be noticed: Firstly,
the z-axis of the
octahedra tend to become aligned with the b axis.
Secondly, the chains dimerize further.

\noindent

\begin{figure}
    \includegraphics[width=\linewidth]{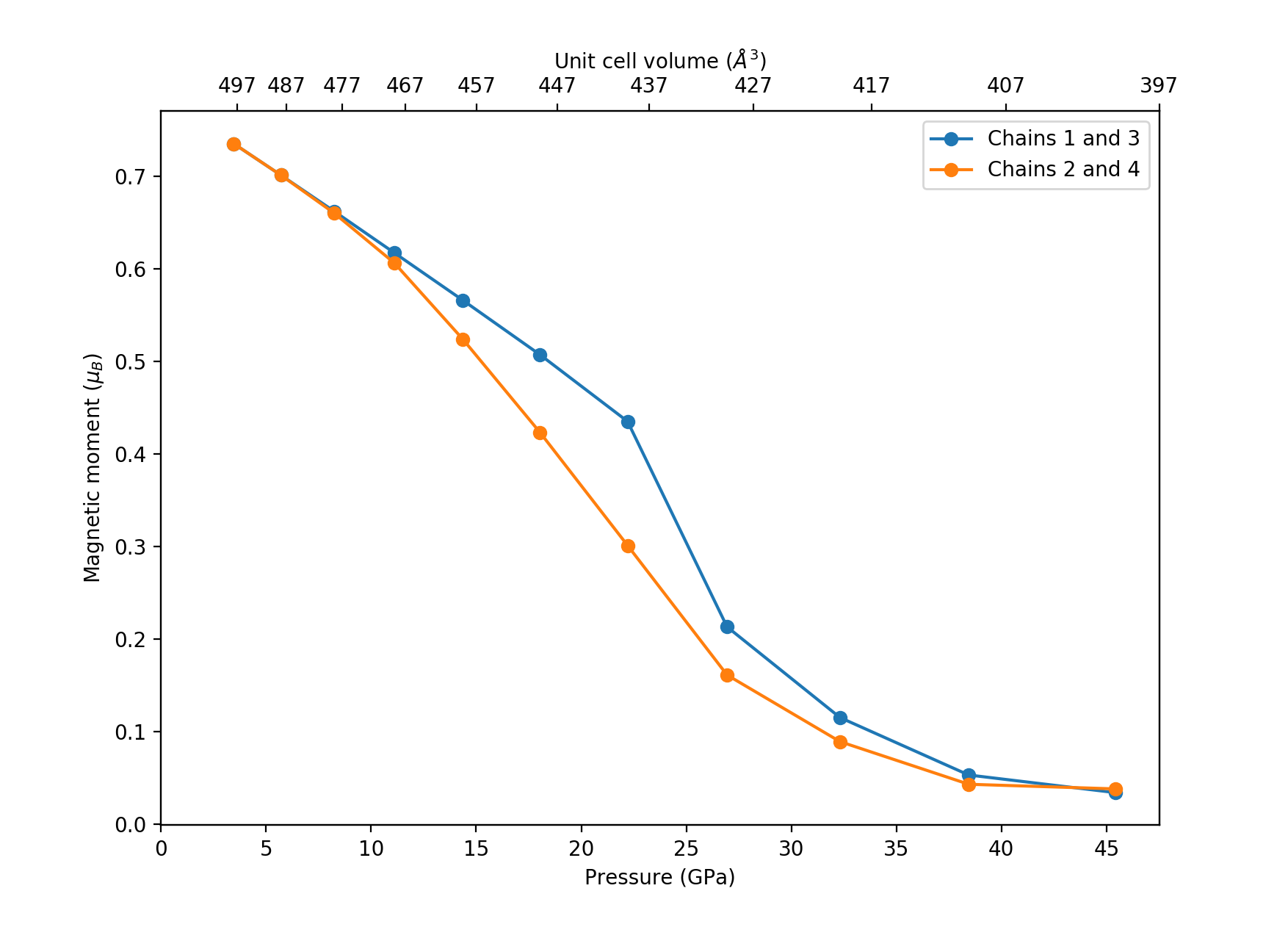}
    \caption{Calculated magnetic moments of Ti atoms, obtained in simulations of
    phase III with induced AFM order using the LDA+U method, as a
function of pressure. Unit cell volume is shown on the upper x-axis.}
    \label{fig:magmom}
\end{figure}

However, the pressure dependence of the tilt angles is different in the four
different chains. In agreement with experiment, we
find chains 2 and 3 to display a larger tilt than chains 1 and 4, Fig. \ref{fig:angles}. At the same
time, the Ti-Ti distances, as well as local magnetic moments of Ti atoms in chains 2 and 4
are equal to each other, as are those of chains 1 and 3, see Figs.
\ref{fig:dim-AFM} and \ref{fig:magmom}.  The former two chains dimerize at a quicker rate
with pressure, and have smaller magnetic moments.  It thus appears that the
dimerization is negatively correlated with the magnitude of the local magnetic
moments.

\begin{figure}
    \includegraphics[width=\linewidth]{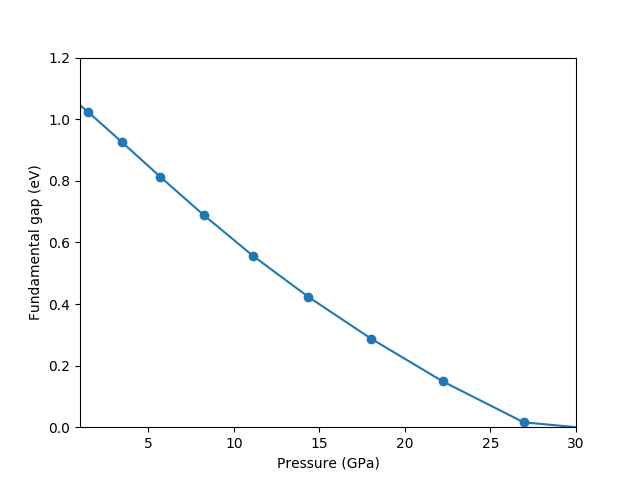}
    \caption{Calculated fundamental gap as a function of pressure.}
    \label{fig:bandgap-P}
\end{figure}

The local Ti DOS depicted in Fig. \ref{fig:Ti-dos} shows that as pressure is increased
and the occupied bands broaden, spin up electrons are transferred to the spin down channel,
leading to a gradual quenching of magnetic moments. As shown in Fig.
\ref{fig:bandgap-P} the insulating gap is also
gradually reduced, and closes completely at a pressure of 28 GPa. Above this point, the system
is metallic in our calculations. As the system is metallized, the magnetic
moments of Ti atoms in chains 1 and 3 drop sharply
and they become similar to those of chains 2 and 4. At the same time, chains 1 and 3
drastically increase their dimerization and become similar to chains 2 and 4
also in this respect. These effects are also visible in the orientation of the
octahedra. Chain 3 is tilted and becomes similar to chain 2, while chain 1 is
straightened out to become similar to chain 4, being nearly aligned with the b-direction.
In the metallic regime the chains thus have the same interatomic distances and
magnetic moments, but the unit cell is characterized by two distinguishable
chains with different tilt angles. Both chains are dimerized,
but one is more tilted than the other. Increasing pressure further in the
metallic regime, we find that the
dimerization start to decrease in both chains, although they retain a
significant degree of dimerization.
The high-pressure phase of TiPO$_4$ is thus strongly reminiscent of
the so-called $M_2$ phase of VO$_2$, which can be induced by uniaxial
pressure or minute amounts of Cr-doping.

It appears that the metallization is directly connected with dimerization of
the chains. By freezing the internal coordinates of the atoms in the unit cell,
and rescaling the lattice
constants according to Fig. \ref{fig:normLat}, we find that the bands broaden
much slower and that the magnitude of the gap remains the same even up to the highest pressure
 considered in our calculations(see Fig. \ref{fig:tot-dos}).

 It is important to point out that there is a subtlety concerning the choice of
 symmetries -- non-magnetic
 or antiferromagnetic -- in the calculations.
 Experimentally, the TiPO$_4$ phase I has been suggested to form a
 singlet dimer phase
 at low temperatures \cite{jrn:law11}.
 There is thus no magnetic order. Nevertheless we choose in the
 calculations to allow for antiferromagnetic ordering. Indeed,
 we argue that the energetics of a dimerized phase
 can be expected to be qualitatively and semi-quantitatively
 well described by a mean-field calculation
 when (artificially) allowing for antiferromagnetic ordering.

 The reason can be seen on the simple example of the Hubbard dimer
 \cite{phd:tomczak}.
 At half filling the ground state of a Hubbard dimer defined
 by hopping $t$ and Hubbard interaction $U$ is a spin singlet, and
 its energy reads:
 \begin{eqnarray}
   E_0 = \frac{U}{2} - \frac{1}{2} \sqrt{U^2 + (4t)^2}
 \end{eqnarray}
 At large interaction $U>>t$, $E_0 = - \frac{4t^2}{U}$ corresponds
 to the spin fluctuation energy gain.
 A mean field theory preserving the spin singlet symmetry
 (unrestricted Hartree-Fock) yields the energy $\frac{U}{2} - 2t$,
 while an antiferromagnetic mean field solution yields
 $- \frac{2t^2}{U}$. While the slope of the energy gain with $t$
 is thus wrong by a factor of $2$, it is clear that the antiferromagnetic
 solution describes the energetics of the system at least semi-quantitatively,
 while the paramagnetic solution does not.

 The above discussion is a special case of the well-known symmetry
 dilemma of density functional theory \cite{jrn:perdew95},
 stating that -- when approximate functional are used -- 
 Kohn-Sham solutions that do not respect the symmetry
 of the physical ground state but rather introduce artificial
 symmetry breakings can yield better energetics than Kohn-Sham
 solutions with the correct symmetry of the physical ground state.
 For this reason, in our calculations, we allow for antiferromagnetic
 order of the magnetic moments along the chains in the c-direction
 of the considered supercells. Of course, further experiments as well as
 theoretical studies, e.g. cluster DMFT, are highly desireable to clarify the
 nature of magnetism in phase III.

\noindent

\noindent

\section{\label{sec:Conc}Summary and conclusions}\noindent
We have investigated the evolution of the crystal structure of TiPO4 from ambient to high pressure,
considering phases I and III, by means of theoretical calculations employing
the LDA+U method. Though at ambient pressure
we do not detect any dimerization of the Ti atoms chains, similar to the
spin-Peierls compound TiOCl \cite{jrn:pisani07}, as pressure increases, we find
that the chains start to dimerize.
The four Ti chains react differently to pressure in the insulating phase but
become equally dimerized as the system is metallized. Remarkably, the
dimerization persists even in the metallic regime, where the magnetic
moments of the Ti atoms are very small. Moreover, if the internal coordinates of
atoms in the simulation cell are kept fixed to
their undimerized positions and the lattice constants are simply rescaled, we
find the metallization to occur at a much higher pressure (smaller volume).

Our results thus indicate that the insulator-to-metal transition is interlinked
with the dimerization of the chains at high pressure. The gradual loss of
magnetism upon compression observed in our calculations seems to increase the
dimerization of the chains. Although our calculations quantitatively underestimate
the degree of dimerization, they capture a picture of the
influence of pressure on the  crystal structure qualitatively correctly. This
adds an interesting new aspect to the
physics of the quasi-one dimensional Ti chains in TiPO$_4$, the metallization
must be considered as an important part of the process.

We
suggest that TiPO$_4$ represents an interesting new model system for studies of
the interplay between the Mott insulator-to-metal transition and the spin-Peierls
effect, and that further studies on TiPO$_4$ should include non-local correlation
effects, which are known to be important in both TiOCl and VO$_2$.

\section*{Acknowledgements}\noindent
This project is funded by the Knut and Alice Wallenberg Foundation
(Wallenberg Scholar grant No. KAW-2018.0194).
We are grateful to the Swedish e-Science Research Centre (SERC) for financial support.
I.A.A. gratefully acknowledges the Swedish Research Council (VR) grant No.
2015-04391 and the Swedish Government Strategic Research Area in Materials
Science on Functional Materials at Linköping University (Faculty Grant
SFO-Mat-LiU No. 2009 00971).
The computations were performed on resources provided by the Swedish National
Infrastructure for Computing (SNIC) at High Performance Computing Center North
(HPC2N) and National Supercomputer Centre (NSC).
We further acknowledge support by the European Research Council
(Project CorrelMat 617196) and IDRIS GENCI Orsay (project 0901393)
at the beginning of the project.

\bibliographystyle{apsrev4-2}
\bibliography{p-iii}

\begin{thebibliography}{36}%
\makeatletter
\providecommand \@ifxundefined [1]{%
 \@ifx{#1\undefined}
}%
\providecommand \@ifnum [1]{%
 \ifnum #1\expandafter \@firstoftwo
 \else \expandafter \@secondoftwo
 \fi
}%
\providecommand \@ifx [1]{%
 \ifx #1\expandafter \@firstoftwo
 \else \expandafter \@secondoftwo
 \fi
}%
\providecommand \natexlab [1]{#1}%
\providecommand \enquote  [1]{``#1''}%
\providecommand \bibnamefont  [1]{#1}%
\providecommand \bibfnamefont [1]{#1}%
\providecommand \citenamefont [1]{#1}%
\providecommand \href@noop [0]{\@secondoftwo}%
\providecommand \href [0]{\begingroup \@sanitize@url \@href}%
\providecommand \@href[1]{\@@startlink{#1}\@@href}%
\providecommand \@@href[1]{\endgroup#1\@@endlink}%
\providecommand \@sanitize@url [0]{\catcode `\\12\catcode `\$12\catcode
  `\&12\catcode `\#12\catcode `\^12\catcode `\_12\catcode `\%12\relax}%
\providecommand \@@startlink[1]{}%
\providecommand \@@endlink[0]{}%
\providecommand \url  [0]{\begingroup\@sanitize@url \@url }%
\providecommand \@url [1]{\endgroup\@href {#1}{\urlprefix }}%
\providecommand \urlprefix  [0]{URL }%
\providecommand \Eprint [0]{\href }%
\providecommand \doibase [0]{https://doi.org/}%
\providecommand \selectlanguage [0]{\@gobble}%
\providecommand \bibinfo  [0]{\@secondoftwo}%
\providecommand \bibfield  [0]{\@secondoftwo}%
\providecommand \translation [1]{[#1]}%
\providecommand \BibitemOpen [0]{}%
\providecommand \bibitemStop [0]{}%
\providecommand \bibitemNoStop [0]{.\EOS\space}%
\providecommand \EOS [0]{\spacefactor3000\relax}%
\providecommand \BibitemShut  [1]{\csname bibitem#1\endcsname}%
\let\auto@bib@innerbib\@empty
\bibitem [{\citenamefont {Dagotto}(2005)}]{jrn:dagotto}%
  \BibitemOpen
  \bibfield  {author} {\bibinfo {author} {\bibfnamefont {E.}~\bibnamefont
  {Dagotto}},\ }\href {https://doi.org/10.1126/science.1107559} {\bibfield
  {journal} {\bibinfo  {journal} {Science}\ }\textbf {\bibinfo {volume}
  {309}},\ \bibinfo {pages} {257} (\bibinfo {year} {2005})},\ \Eprint
  {https://arxiv.org/abs/http://science.sciencemag.org/content/309/5732/257.full.pdf}
  {http://science.sciencemag.org/content/309/5732/257.full.pdf} \BibitemShut
  {NoStop}%
\bibitem [{\citenamefont {Stepanov}\ \emph {et~al.}(2018)\citenamefont
  {Stepanov}, \citenamefont {Peters}, \citenamefont {Krivenko}, \citenamefont
  {Lichtenstein}, \citenamefont {Katsnelson},\ and\ \citenamefont
  {Rubtsov}}]{jrn:stepanov}%
  \BibitemOpen
  \bibfield  {author} {\bibinfo {author} {\bibfnamefont {E.~A.}\ \bibnamefont
  {Stepanov}}, \bibinfo {author} {\bibfnamefont {L.}~\bibnamefont {Peters}},
  \bibinfo {author} {\bibfnamefont {I.~S.}\ \bibnamefont {Krivenko}}, \bibinfo
  {author} {\bibfnamefont {A.~I.}\ \bibnamefont {Lichtenstein}}, \bibinfo
  {author} {\bibfnamefont {M.~I.}\ \bibnamefont {Katsnelson}},\ and\ \bibinfo
  {author} {\bibfnamefont {A.~N.}\ \bibnamefont {Rubtsov}},\ }\href@noop {}
  {\bibfield  {journal} {\bibinfo  {journal} {npj Quantum Materials}\ }\textbf
  {\bibinfo {volume} {3}},\ \bibinfo {pages} {54} (\bibinfo {year}
  {2018})}\BibitemShut {NoStop}%
\bibitem [{\citenamefont {Hase}\ \emph {et~al.}(1993)\citenamefont {Hase},
  \citenamefont {Terasaki}, \citenamefont {Uchinokura}, \citenamefont
  {Tokunaga}, \citenamefont {Miura},\ and\ \citenamefont {Obara}}]{jrn:hase93}%
  \BibitemOpen
  \bibfield  {author} {\bibinfo {author} {\bibfnamefont {M.}~\bibnamefont
  {Hase}}, \bibinfo {author} {\bibfnamefont {I.}~\bibnamefont {Terasaki}},
  \bibinfo {author} {\bibfnamefont {K.}~\bibnamefont {Uchinokura}}, \bibinfo
  {author} {\bibfnamefont {M.}~\bibnamefont {Tokunaga}}, \bibinfo {author}
  {\bibfnamefont {N.}~\bibnamefont {Miura}},\ and\ \bibinfo {author}
  {\bibfnamefont {H.}~\bibnamefont {Obara}},\ }\href
  {https://doi.org/10.1103/PhysRevB.48.9616} {\bibfield  {journal} {\bibinfo
  {journal} {Phys. Rev. B}\ }\textbf {\bibinfo {volume} {48}},\ \bibinfo
  {pages} {9616} (\bibinfo {year} {1993})}\BibitemShut {NoStop}%
\bibitem [{\citenamefont {Seidel}\ \emph {et~al.}(2003)\citenamefont {Seidel},
  \citenamefont {Marianetti}, \citenamefont {Chou}, \citenamefont {Ceder},\
  and\ \citenamefont {Lee}}]{jrn:seidel03}%
  \BibitemOpen
  \bibfield  {author} {\bibinfo {author} {\bibfnamefont {A.}~\bibnamefont
  {Seidel}}, \bibinfo {author} {\bibfnamefont {C.~A.}\ \bibnamefont
  {Marianetti}}, \bibinfo {author} {\bibfnamefont {F.~C.}\ \bibnamefont
  {Chou}}, \bibinfo {author} {\bibfnamefont {G.}~\bibnamefont {Ceder}},\ and\
  \bibinfo {author} {\bibfnamefont {P.~A.}\ \bibnamefont {Lee}},\ }\href
  {https://doi.org/10.1103/PhysRevB.67.020405} {\bibfield  {journal} {\bibinfo
  {journal} {Phys. Rev. B}\ }\textbf {\bibinfo {volume} {67}},\ \bibinfo
  {pages} {020405} (\bibinfo {year} {2003})}\BibitemShut {NoStop}%
\bibitem [{\citenamefont {Shaz}\ \emph {et~al.}(2005)\citenamefont {Shaz},
  \citenamefont {van Smaalen}, \citenamefont {Palatinus}, \citenamefont
  {Hoinkis}, \citenamefont {Klemm}, \citenamefont {Horn},\ and\ \citenamefont
  {Claessen}}]{jrn:shaz05}%
  \BibitemOpen
  \bibfield  {author} {\bibinfo {author} {\bibfnamefont {M.}~\bibnamefont
  {Shaz}}, \bibinfo {author} {\bibfnamefont {S.}~\bibnamefont {van Smaalen}},
  \bibinfo {author} {\bibfnamefont {L.}~\bibnamefont {Palatinus}}, \bibinfo
  {author} {\bibfnamefont {M.}~\bibnamefont {Hoinkis}}, \bibinfo {author}
  {\bibfnamefont {M.}~\bibnamefont {Klemm}}, \bibinfo {author} {\bibfnamefont
  {S.}~\bibnamefont {Horn}},\ and\ \bibinfo {author} {\bibfnamefont
  {R.}~\bibnamefont {Claessen}},\ }\href
  {https://doi.org/10.1103/PhysRevB.71.100405} {\bibfield  {journal} {\bibinfo
  {journal} {Phys. Rev. B}\ }\textbf {\bibinfo {volume} {71}},\ \bibinfo
  {pages} {100405} (\bibinfo {year} {2005})}\BibitemShut {NoStop}%
\bibitem [{\citenamefont {Saha-Dasgupta}\ \emph {et~al.}(2005)\citenamefont
  {Saha-Dasgupta}, \citenamefont {Lichtenstein},\ and\ \citenamefont
  {Valent\'{\i}}}]{jrn:sahadasgupta05}%
  \BibitemOpen
  \bibfield  {author} {\bibinfo {author} {\bibfnamefont {T.}~\bibnamefont
  {Saha-Dasgupta}}, \bibinfo {author} {\bibfnamefont {A.}~\bibnamefont
  {Lichtenstein}},\ and\ \bibinfo {author} {\bibfnamefont {R.}~\bibnamefont
  {Valent\'{\i}}},\ }\href {https://doi.org/10.1103/PhysRevB.71.153108}
  {\bibfield  {journal} {\bibinfo  {journal} {Phys. Rev. B}\ }\textbf {\bibinfo
  {volume} {71}},\ \bibinfo {pages} {153108} (\bibinfo {year}
  {2005})}\BibitemShut {NoStop}%
\bibitem [{\citenamefont {Saha-Dasgupta}\ \emph {et~al.}(2007)\citenamefont
  {Saha-Dasgupta}, \citenamefont {Lichtenstein}, \citenamefont {Hoinkis},
  \citenamefont {Glawion}, \citenamefont {Sing}, \citenamefont {Claessen},\
  and\ \citenamefont {Valent{\'{\i}}}}]{jrn:sahadasgupta07}%
  \BibitemOpen
  \bibfield  {author} {\bibinfo {author} {\bibfnamefont {T.}~\bibnamefont
  {Saha-Dasgupta}}, \bibinfo {author} {\bibfnamefont {A.}~\bibnamefont
  {Lichtenstein}}, \bibinfo {author} {\bibfnamefont {M.}~\bibnamefont
  {Hoinkis}}, \bibinfo {author} {\bibfnamefont {S.}~\bibnamefont {Glawion}},
  \bibinfo {author} {\bibfnamefont {M.}~\bibnamefont {Sing}}, \bibinfo {author}
  {\bibfnamefont {R.}~\bibnamefont {Claessen}},\ and\ \bibinfo {author}
  {\bibfnamefont {R.}~\bibnamefont {Valent{\'{\i}}}},\ }\href
  {https://doi.org/10.1088/1367-2630/9/10/380} {\bibfield  {journal} {\bibinfo
  {journal} {New Journal of Physics}\ }\textbf {\bibinfo {volume} {9}},\
  \bibinfo {pages} {380} (\bibinfo {year} {2007})}\BibitemShut {NoStop}%
\bibitem [{\citenamefont {Aichhorn}\ \emph {et~al.}(2009)\citenamefont
  {Aichhorn}, \citenamefont {Saha-Dasgupta}, \citenamefont {Valent\'{\i}},
  \citenamefont {Glawion}, \citenamefont {Sing},\ and\ \citenamefont
  {Claessen}}]{jrn:aichhorn09}%
  \BibitemOpen
  \bibfield  {author} {\bibinfo {author} {\bibfnamefont {M.}~\bibnamefont
  {Aichhorn}}, \bibinfo {author} {\bibfnamefont {T.}~\bibnamefont
  {Saha-Dasgupta}}, \bibinfo {author} {\bibfnamefont {R.}~\bibnamefont
  {Valent\'{\i}}}, \bibinfo {author} {\bibfnamefont {S.}~\bibnamefont
  {Glawion}}, \bibinfo {author} {\bibfnamefont {M.}~\bibnamefont {Sing}},\ and\
  \bibinfo {author} {\bibfnamefont {R.}~\bibnamefont {Claessen}},\ }\href
  {https://doi.org/10.1103/PhysRevB.80.115129} {\bibfield  {journal} {\bibinfo
  {journal} {Phys. Rev. B}\ }\textbf {\bibinfo {volume} {80}},\ \bibinfo
  {pages} {115129} (\bibinfo {year} {2009})}\BibitemShut {NoStop}%
\bibitem [{\citenamefont {Rotundu}\ \emph {et~al.}(2018)\citenamefont
  {Rotundu}, \citenamefont {Wen}, \citenamefont {He}, \citenamefont {Choi},
  \citenamefont {Haskel},\ and\ \citenamefont {Lee}}]{jrn:rotundu18}%
  \BibitemOpen
  \bibfield  {author} {\bibinfo {author} {\bibfnamefont {C.~R.}\ \bibnamefont
  {Rotundu}}, \bibinfo {author} {\bibfnamefont {J.}~\bibnamefont {Wen}},
  \bibinfo {author} {\bibfnamefont {W.}~\bibnamefont {He}}, \bibinfo {author}
  {\bibfnamefont {Y.}~\bibnamefont {Choi}}, \bibinfo {author} {\bibfnamefont
  {D.}~\bibnamefont {Haskel}},\ and\ \bibinfo {author} {\bibfnamefont {Y.~S.}\
  \bibnamefont {Lee}},\ }\href {https://doi.org/10.1103/PhysRevB.97.054415}
  {\bibfield  {journal} {\bibinfo  {journal} {Phys. Rev. B}\ }\textbf {\bibinfo
  {volume} {97}},\ \bibinfo {pages} {054415} (\bibinfo {year}
  {2018})}\BibitemShut {NoStop}%
\bibitem [{\citenamefont {Bykov}\ \emph {et~al.}(2016)\citenamefont {Bykov}
  \emph {et~al.}}]{jrn:bykov16}%
  \BibitemOpen
  \bibfield  {author} {\bibinfo {author} {\bibfnamefont {M.}~\bibnamefont
  {Bykov}} \emph {et~al.},\ }\bibfield  {journal} {\bibinfo  {journal}
  {Angewandte Chemie International Edition}\ }\textbf {\bibinfo {volume}
  {55}},\ \href {https://doi.org/10.1002/anie.201608530}
  {10.1002/anie.201608530} (\bibinfo {year} {2016})\BibitemShut {NoStop}%
\bibitem [{\citenamefont {Kinomura}\ \emph {et~al.}(1982)\citenamefont
  {Kinomura}, \citenamefont {Muto},\ and\ \citenamefont
  {Koizumi}}]{jrn:kinomura82}%
  \BibitemOpen
  \bibfield  {author} {\bibinfo {author} {\bibfnamefont {N.}~\bibnamefont
  {Kinomura}}, \bibinfo {author} {\bibfnamefont {F.}~\bibnamefont {Muto}},\
  and\ \bibinfo {author} {\bibfnamefont {M.}~\bibnamefont {Koizumi}},\ }\href
  {https://doi.org/\url{https://doi.org/10.1016/0022-4596(82)90281-X}}
  {\bibfield  {journal} {\bibinfo  {journal} {Journal of Solid State
  Chemistry}\ }\textbf {\bibinfo {volume} {45}},\ \bibinfo {pages} {252 }
  (\bibinfo {year} {1982})}\BibitemShut {NoStop}%
\bibitem [{\citenamefont {Glaum}\ \emph {et~al.}(1996)\citenamefont {Glaum},
  \citenamefont {Reehuis}, \citenamefont {Stüßer}, \citenamefont {Kaiser},\
  and\ \citenamefont {Reinauer}}]{jrn:glaum96}%
  \BibitemOpen
  \bibfield  {author} {\bibinfo {author} {\bibfnamefont {R.}~\bibnamefont
  {Glaum}}, \bibinfo {author} {\bibfnamefont {M.}~\bibnamefont {Reehuis}},
  \bibinfo {author} {\bibfnamefont {N.}~\bibnamefont {Stüßer}}, \bibinfo
  {author} {\bibfnamefont {U.}~\bibnamefont {Kaiser}},\ and\ \bibinfo {author}
  {\bibfnamefont {F.}~\bibnamefont {Reinauer}},\ }\href
  {https://doi.org/https://doi.org/10.1006/jssc.1996.0303} {\bibfield
  {journal} {\bibinfo  {journal} {Journal of Solid State Chemistry}\ }\textbf
  {\bibinfo {volume} {126}},\ \bibinfo {pages} {15 } (\bibinfo {year}
  {1996})}\BibitemShut {NoStop}%
\bibitem [{\citenamefont {Law}\ \emph {et~al.}(2011)\citenamefont {Law},
  \citenamefont {Hoch}, \citenamefont {Glaum}, \citenamefont {Heinmaa},
  \citenamefont {Stern}, \citenamefont {Kang}, \citenamefont {Lee},
  \citenamefont {Whangbo},\ and\ \citenamefont {Kremer}}]{jrn:law11}%
  \BibitemOpen
  \bibfield  {author} {\bibinfo {author} {\bibfnamefont {J.~M.}\ \bibnamefont
  {Law}}, \bibinfo {author} {\bibfnamefont {C.}~\bibnamefont {Hoch}}, \bibinfo
  {author} {\bibfnamefont {R.}~\bibnamefont {Glaum}}, \bibinfo {author}
  {\bibfnamefont {I.}~\bibnamefont {Heinmaa}}, \bibinfo {author} {\bibfnamefont
  {R.}~\bibnamefont {Stern}}, \bibinfo {author} {\bibfnamefont
  {J.}~\bibnamefont {Kang}}, \bibinfo {author} {\bibfnamefont {C.}~\bibnamefont
  {Lee}}, \bibinfo {author} {\bibfnamefont {M.-H.}\ \bibnamefont {Whangbo}},\
  and\ \bibinfo {author} {\bibfnamefont {R.~K.}\ \bibnamefont {Kremer}},\
  }\href {https://doi.org/10.1103/PhysRevB.83.180414} {\bibfield  {journal}
  {\bibinfo  {journal} {Phys. Rev. B}\ }\textbf {\bibinfo {volume} {83}},\
  \bibinfo {pages} {180414} (\bibinfo {year} {2011})}\BibitemShut {NoStop}%
\bibitem [{\citenamefont {Bykov}\ \emph {et~al.}(2013)\citenamefont {Bykov}
  \emph {et~al.}}]{jrn:bykov13}%
  \BibitemOpen
  \bibfield  {author} {\bibinfo {author} {\bibfnamefont {M.}~\bibnamefont
  {Bykov}} \emph {et~al.},\ }\bibfield  {journal} {\bibinfo  {journal} {Phys.
  Rev. B}\ }\textbf {\bibinfo {volume} {88}},\ \href
  {https://doi.org/10.1103/PhysRevB.88.184420} {10.1103/PhysRevB.88.184420}
  (\bibinfo {year} {2013})\BibitemShut {NoStop}%
\bibitem [{\citenamefont {MOTT}(1968)}]{jrn:mottreview68}%
  \BibitemOpen
  \bibfield  {author} {\bibinfo {author} {\bibfnamefont {N.~F.}\ \bibnamefont
  {MOTT}},\ }\href {https://doi.org/10.1103/RevModPhys.40.677} {\bibfield
  {journal} {\bibinfo  {journal} {Rev. Mod. Phys.}\ }\textbf {\bibinfo {volume}
  {40}},\ \bibinfo {pages} {677} (\bibinfo {year} {1968})}\BibitemShut
  {NoStop}%
\bibitem [{\citenamefont {Goodenough}(1971)}]{jrn:goodenough71}%
  \BibitemOpen
  \bibfield  {author} {\bibinfo {author} {\bibfnamefont {J.~B.}\ \bibnamefont
  {Goodenough}},\ }\href
  {https://doi.org/https://doi.org/10.1016/0022-4596(71)90091-0} {\bibfield
  {journal} {\bibinfo  {journal} {Journal of Solid State Chemistry}\ }\textbf
  {\bibinfo {volume} {3}},\ \bibinfo {pages} {490 } (\bibinfo {year}
  {1971})}\BibitemShut {NoStop}%
\bibitem [{\citenamefont {Eguchi}\ \emph {et~al.}(2008)\citenamefont {Eguchi},
  \citenamefont {Taguchi}, \citenamefont {Matsunami}, \citenamefont {Horiba},
  \citenamefont {Yamamoto}, \citenamefont {Ishida}, \citenamefont {Chainani},
  \citenamefont {Takata}, \citenamefont {Yabashi}, \citenamefont {Miwa},
  \citenamefont {Nishino}, \citenamefont {Tamasaku}, \citenamefont {Ishikawa},
  \citenamefont {Senba}, \citenamefont {Ohashi}, \citenamefont {Muraoka},
  \citenamefont {Hiroi},\ and\ \citenamefont {Shin}}]{jrn:eguchi08}%
  \BibitemOpen
  \bibfield  {author} {\bibinfo {author} {\bibfnamefont {R.}~\bibnamefont
  {Eguchi}}, \bibinfo {author} {\bibfnamefont {M.}~\bibnamefont {Taguchi}},
  \bibinfo {author} {\bibfnamefont {M.}~\bibnamefont {Matsunami}}, \bibinfo
  {author} {\bibfnamefont {K.}~\bibnamefont {Horiba}}, \bibinfo {author}
  {\bibfnamefont {K.}~\bibnamefont {Yamamoto}}, \bibinfo {author}
  {\bibfnamefont {Y.}~\bibnamefont {Ishida}}, \bibinfo {author} {\bibfnamefont
  {A.}~\bibnamefont {Chainani}}, \bibinfo {author} {\bibfnamefont
  {Y.}~\bibnamefont {Takata}}, \bibinfo {author} {\bibfnamefont
  {M.}~\bibnamefont {Yabashi}}, \bibinfo {author} {\bibfnamefont
  {D.}~\bibnamefont {Miwa}}, \bibinfo {author} {\bibfnamefont {Y.}~\bibnamefont
  {Nishino}}, \bibinfo {author} {\bibfnamefont {K.}~\bibnamefont {Tamasaku}},
  \bibinfo {author} {\bibfnamefont {T.}~\bibnamefont {Ishikawa}}, \bibinfo
  {author} {\bibfnamefont {Y.}~\bibnamefont {Senba}}, \bibinfo {author}
  {\bibfnamefont {H.}~\bibnamefont {Ohashi}}, \bibinfo {author} {\bibfnamefont
  {Y.}~\bibnamefont {Muraoka}}, \bibinfo {author} {\bibfnamefont
  {Z.}~\bibnamefont {Hiroi}},\ and\ \bibinfo {author} {\bibfnamefont
  {S.}~\bibnamefont {Shin}},\ }\href
  {https://doi.org/10.1103/PhysRevB.78.075115} {\bibfield  {journal} {\bibinfo
  {journal} {Phys. Rev. B}\ }\textbf {\bibinfo {volume} {78}},\ \bibinfo
  {pages} {075115} (\bibinfo {year} {2008})}\BibitemShut {NoStop}%
\bibitem [{\citenamefont {Eyert}(2002)}]{jrn:eyert02}%
  \BibitemOpen
  \bibfield  {author} {\bibinfo {author} {\bibfnamefont {V.}~\bibnamefont
  {Eyert}},\ }\href
  {https://doi.org/10.1002/1521-3889(200210)11:9<650::AID-ANDP650>3.0.CO;2-K}
  {\bibfield  {journal} {\bibinfo  {journal} {Annalen der Physik}\ }\textbf
  {\bibinfo {volume} {11}},\ \bibinfo {pages} {650} (\bibinfo {year}
  {2002})}\BibitemShut {NoStop}%
\bibitem [{\citenamefont {Biermann}\ \emph {et~al.}(2005)\citenamefont
  {Biermann}, \citenamefont {Poteryaev}, \citenamefont {Lichtenstein},\ and\
  \citenamefont {Georges}}]{jrn:biermann05}%
  \BibitemOpen
  \bibfield  {author} {\bibinfo {author} {\bibfnamefont {S.}~\bibnamefont
  {Biermann}}, \bibinfo {author} {\bibfnamefont {A.}~\bibnamefont {Poteryaev}},
  \bibinfo {author} {\bibfnamefont {A.~I.}\ \bibnamefont {Lichtenstein}},\ and\
  \bibinfo {author} {\bibfnamefont {A.}~\bibnamefont {Georges}},\ }\href
  {https://doi.org/10.1103/PhysRevLett.94.026404} {\bibfield  {journal}
  {\bibinfo  {journal} {Phys. Rev. Lett.}\ }\textbf {\bibinfo {volume} {94}},\
  \bibinfo {pages} {026404} (\bibinfo {year} {2005})}\BibitemShut {NoStop}%
\bibitem [{\citenamefont {Tomczak}\ \emph {et~al.}(2008)\citenamefont
  {Tomczak}, \citenamefont {Aryasetiawan},\ and\ \citenamefont
  {Biermann}}]{jrn:tomczak08}%
  \BibitemOpen
  \bibfield  {author} {\bibinfo {author} {\bibfnamefont {J.~M.}\ \bibnamefont
  {Tomczak}}, \bibinfo {author} {\bibfnamefont {F.}~\bibnamefont
  {Aryasetiawan}},\ and\ \bibinfo {author} {\bibfnamefont {S.}~\bibnamefont
  {Biermann}},\ }\href {https://doi.org/10.1103/PhysRevB.78.115103} {\bibfield
  {journal} {\bibinfo  {journal} {Phys. Rev. B}\ }\textbf {\bibinfo {volume}
  {78}},\ \bibinfo {pages} {115103} (\bibinfo {year} {2008})}\BibitemShut
  {NoStop}%
\bibitem [{\citenamefont {{POUGET, J. P.}}\ and\ \citenamefont {{LAUNOIS,
  H.}}(1976)}]{jrn:pouget76}%
  \BibitemOpen
  \bibfield  {author} {\bibinfo {author} {\bibnamefont {{POUGET, J. P.}}}\ and\
  \bibinfo {author} {\bibnamefont {{LAUNOIS, H.}}},\ }\href
  {https://doi.org/10.1051/jphyscol:1976408} {\bibfield  {journal} {\bibinfo
  {journal} {J. Phys. Colloques}\ }\textbf {\bibinfo {volume} {37}},\ \bibinfo
  {pages} {C4} (\bibinfo {year} {1976})}\BibitemShut {NoStop}%
\bibitem [{\citenamefont {Koethe}\ \emph {et~al.}(2006)\citenamefont {Koethe},
  \citenamefont {Hu}, \citenamefont {Haverkort}, \citenamefont
  {Sch\"u\ss{}ler-Langeheine}, \citenamefont {Venturini}, \citenamefont
  {Brookes}, \citenamefont {Tjernberg}, \citenamefont {Reichelt}, \citenamefont
  {Hsieh}, \citenamefont {Lin}, \citenamefont {Chen},\ and\ \citenamefont
  {Tjeng}}]{jrn:koethe06}%
  \BibitemOpen
  \bibfield  {author} {\bibinfo {author} {\bibfnamefont {T.~C.}\ \bibnamefont
  {Koethe}}, \bibinfo {author} {\bibfnamefont {Z.}~\bibnamefont {Hu}}, \bibinfo
  {author} {\bibfnamefont {M.~W.}\ \bibnamefont {Haverkort}}, \bibinfo {author}
  {\bibfnamefont {C.}~\bibnamefont {Sch\"u\ss{}ler-Langeheine}}, \bibinfo
  {author} {\bibfnamefont {F.}~\bibnamefont {Venturini}}, \bibinfo {author}
  {\bibfnamefont {N.~B.}\ \bibnamefont {Brookes}}, \bibinfo {author}
  {\bibfnamefont {O.}~\bibnamefont {Tjernberg}}, \bibinfo {author}
  {\bibfnamefont {W.}~\bibnamefont {Reichelt}}, \bibinfo {author}
  {\bibfnamefont {H.~H.}\ \bibnamefont {Hsieh}}, \bibinfo {author}
  {\bibfnamefont {H.-J.}\ \bibnamefont {Lin}}, \bibinfo {author} {\bibfnamefont
  {C.~T.}\ \bibnamefont {Chen}},\ and\ \bibinfo {author} {\bibfnamefont
  {L.~H.}\ \bibnamefont {Tjeng}},\ }\href
  {https://doi.org/10.1103/PhysRevLett.97.116402} {\bibfield  {journal}
  {\bibinfo  {journal} {Phys. Rev. Lett.}\ }\textbf {\bibinfo {volume} {97}},\
  \bibinfo {pages} {116402} (\bibinfo {year} {2006})}\BibitemShut {NoStop}%
\bibitem [{\citenamefont {Tan}\ \emph {et~al.}(2012)\citenamefont {Tan},
  \citenamefont {Yao}, \citenamefont {Long}, \citenamefont {Sun}, \citenamefont
  {Feng}, \citenamefont {Cheng}, \citenamefont {Yuan}, \citenamefont {Zhang},
  \citenamefont {Liu}, \citenamefont {Wu}, \citenamefont {Xie},\ and\
  \citenamefont {Wei}}]{jrn:tan12}%
  \BibitemOpen
  \bibfield  {author} {\bibinfo {author} {\bibfnamefont {X.}~\bibnamefont
  {Tan}}, \bibinfo {author} {\bibfnamefont {T.}~\bibnamefont {Yao}}, \bibinfo
  {author} {\bibfnamefont {R.}~\bibnamefont {Long}}, \bibinfo {author}
  {\bibfnamefont {Z.}~\bibnamefont {Sun}}, \bibinfo {author} {\bibfnamefont
  {Y.}~\bibnamefont {Feng}}, \bibinfo {author} {\bibfnamefont {H.}~\bibnamefont
  {Cheng}}, \bibinfo {author} {\bibfnamefont {X.}~\bibnamefont {Yuan}},
  \bibinfo {author} {\bibfnamefont {W.}~\bibnamefont {Zhang}}, \bibinfo
  {author} {\bibfnamefont {Q.}~\bibnamefont {Liu}}, \bibinfo {author}
  {\bibfnamefont {C.}~\bibnamefont {Wu}}, \bibinfo {author} {\bibfnamefont
  {Y.}~\bibnamefont {Xie}},\ and\ \bibinfo {author} {\bibfnamefont
  {S.}~\bibnamefont {Wei}},\ }\href {https://doi.org/10.1038/srep00466}
  {\bibfield  {journal} {\bibinfo  {journal} {Scientific reports}\ }\textbf
  {\bibinfo {volume} {2}},\ \bibinfo {pages} {466} (\bibinfo {year}
  {2012})}\BibitemShut {NoStop}%
\bibitem [{\citenamefont {Jager}\ \emph {et~al.}(2017)\citenamefont {Jager},
  \citenamefont {Ott}, \citenamefont {Kraus}, \citenamefont {Kaplan},
  \citenamefont {Pouse}, \citenamefont {Marvel}, \citenamefont {Haglund},
  \citenamefont {Neumark},\ and\ \citenamefont {Leone}}]{jrn:jager17}%
  \BibitemOpen
  \bibfield  {author} {\bibinfo {author} {\bibfnamefont {M.~F.}\ \bibnamefont
  {Jager}}, \bibinfo {author} {\bibfnamefont {C.}~\bibnamefont {Ott}}, \bibinfo
  {author} {\bibfnamefont {P.~M.}\ \bibnamefont {Kraus}}, \bibinfo {author}
  {\bibfnamefont {C.~J.}\ \bibnamefont {Kaplan}}, \bibinfo {author}
  {\bibfnamefont {W.}~\bibnamefont {Pouse}}, \bibinfo {author} {\bibfnamefont
  {R.~E.}\ \bibnamefont {Marvel}}, \bibinfo {author} {\bibfnamefont {R.~F.}\
  \bibnamefont {Haglund}}, \bibinfo {author} {\bibfnamefont {D.~M.}\
  \bibnamefont {Neumark}},\ and\ \bibinfo {author} {\bibfnamefont {S.~R.}\
  \bibnamefont {Leone}},\ }\href {https://doi.org/10.1073/pnas.1707602114}
  {\bibfield  {journal} {\bibinfo  {journal} {Proceedings of the National
  Academy of Sciences}\ }\textbf {\bibinfo {volume} {114}},\ \bibinfo {pages}
  {9558} (\bibinfo {year} {2017})},\ \Eprint
  {https://arxiv.org/abs/https://www.pnas.org/content/114/36/9558.full.pdf}
  {https://www.pnas.org/content/114/36/9558.full.pdf} \BibitemShut {NoStop}%
\bibitem [{\citenamefont {Fan}\ \emph {et~al.}(2018)\citenamefont {Fan},
  \citenamefont {Wang}, \citenamefont {Wang}, \citenamefont {Zhang},
  \citenamefont {Zhu}, \citenamefont {Meng}, \citenamefont {Wang},
  \citenamefont {Zhang},\ and\ \citenamefont {Zou}}]{jrn:fan18}%
  \BibitemOpen
  \bibfield  {author} {\bibinfo {author} {\bibfnamefont {L.}~\bibnamefont
  {Fan}}, \bibinfo {author} {\bibfnamefont {X.}~\bibnamefont {Wang}}, \bibinfo
  {author} {\bibfnamefont {F.}~\bibnamefont {Wang}}, \bibinfo {author}
  {\bibfnamefont {Q.}~\bibnamefont {Zhang}}, \bibinfo {author} {\bibfnamefont
  {L.}~\bibnamefont {Zhu}}, \bibinfo {author} {\bibfnamefont {Q.}~\bibnamefont
  {Meng}}, \bibinfo {author} {\bibfnamefont {B.}~\bibnamefont {Wang}}, \bibinfo
  {author} {\bibfnamefont {Z.}~\bibnamefont {Zhang}},\ and\ \bibinfo {author}
  {\bibfnamefont {C.}~\bibnamefont {Zou}},\ }\href
  {https://doi.org/10.1039/C8RA03292K} {\bibfield  {journal} {\bibinfo
  {journal} {RSC Adv.}\ }\textbf {\bibinfo {volume} {8}},\ \bibinfo {pages}
  {19151} (\bibinfo {year} {2018})}\BibitemShut {NoStop}%
\bibitem [{\citenamefont {Hohenberg}\ and\ \citenamefont
  {Kohn}(1964)}]{jrn:hohenbergkohn}%
  \BibitemOpen
  \bibfield  {author} {\bibinfo {author} {\bibfnamefont {P.}~\bibnamefont
  {Hohenberg}}\ and\ \bibinfo {author} {\bibfnamefont {W.}~\bibnamefont
  {Kohn}},\ }\href {https://doi.org/10.1103/PhysRev.136.B864} {\bibfield
  {journal} {\bibinfo  {journal} {Phys. Rev.}\ }\textbf {\bibinfo {volume}
  {136}},\ \bibinfo {pages} {B864} (\bibinfo {year} {1964})}\BibitemShut
  {NoStop}%
\bibitem [{\citenamefont {Kohn}\ and\ \citenamefont
  {Sham}(1965)}]{jrn:kohnsham}%
  \BibitemOpen
  \bibfield  {author} {\bibinfo {author} {\bibfnamefont {W.}~\bibnamefont
  {Kohn}}\ and\ \bibinfo {author} {\bibfnamefont {L.~J.}\ \bibnamefont
  {Sham}},\ }\href {https://doi.org/10.1103/PhysRev.140.A1133} {\bibfield
  {journal} {\bibinfo  {journal} {Phys. Rev.}\ }\textbf {\bibinfo {volume}
  {140}},\ \bibinfo {pages} {A1133} (\bibinfo {year} {1965})}\BibitemShut
  {NoStop}%
\bibitem [{\citenamefont {Bl\"ochl}(1994)}]{jrn:blochl94}%
  \BibitemOpen
  \bibfield  {author} {\bibinfo {author} {\bibfnamefont {P.~E.}\ \bibnamefont
  {Bl\"ochl}},\ }\bibfield  {journal} {\bibinfo  {journal} {Phys. Rev. B}\
  }\textbf {\bibinfo {volume} {50}},\ \href
  {https://doi.org/10.1103/PhysRevB.50.17953} {10.1103/PhysRevB.50.17953}
  (\bibinfo {year} {1994})\BibitemShut {NoStop}%
\bibitem [{\citenamefont {Kresse}\ and\ \citenamefont
  {Furthm\"uller}(1996)}]{jrn:VASP1}%
  \BibitemOpen
  \bibfield  {author} {\bibinfo {author} {\bibfnamefont {G.}~\bibnamefont
  {Kresse}}\ and\ \bibinfo {author} {\bibfnamefont {J.}~\bibnamefont
  {Furthm\"uller}},\ }\bibfield  {journal} {\bibinfo  {journal} {Phys. Rev. B}\
  }\textbf {\bibinfo {volume} {54}},\ \href
  {https://doi.org/10.1103/PhysRevB.54.11169} {10.1103/PhysRevB.54.11169}
  (\bibinfo {year} {1996})\BibitemShut {NoStop}%
\bibitem [{\citenamefont {Kresse}\ and\ \citenamefont
  {Furthm{\"u}ller}(1996)}]{jrn:VASP2}%
  \BibitemOpen
  \bibfield  {author} {\bibinfo {author} {\bibfnamefont {G.}~\bibnamefont
  {Kresse}}\ and\ \bibinfo {author} {\bibfnamefont {J.}~\bibnamefont
  {Furthm{\"u}ller}},\ }\bibfield  {journal} {\bibinfo  {journal}
  {Computational Materials Science}\ }\textbf {\bibinfo {volume} {6}},\ \href
  {https://doi.org/https://doi.org/10.1016/0927-0256(96)00008-0}
  {https://doi.org/10.1016/0927-0256(96)00008-0} (\bibinfo {year}
  {1996})\BibitemShut {NoStop}%
\bibitem [{\citenamefont {Dudarev}\ \emph {et~al.}(1998)\citenamefont {Dudarev}
  \emph {et~al.}}]{jrn:lda+u}%
  \BibitemOpen
  \bibfield  {author} {\bibinfo {author} {\bibfnamefont {S.~L.}\ \bibnamefont
  {Dudarev}} \emph {et~al.},\ }\bibfield  {journal} {\bibinfo  {journal} {Phys.
  Rev. B}\ }\textbf {\bibinfo {volume} {57}},\ \href
  {https://doi.org/10.1103/PhysRevB.57.1505} {10.1103/PhysRevB.57.1505}
  (\bibinfo {year} {1998})\BibitemShut {NoStop}%
\bibitem [{sup()}]{suppmat}%
  \BibitemOpen
  \href@noop {} {}\bibinfo {note} {See supplemental materials for details on
  the experimental methods.}\BibitemShut {Stop}%
\bibitem [{\citenamefont {L\'opez-Moreno}\ and\ \citenamefont
  {Errandonea}(2012)}]{jrn:lopezmoreno}%
  \BibitemOpen
  \bibfield  {author} {\bibinfo {author} {\bibfnamefont {S.}~\bibnamefont
  {L\'opez-Moreno}}\ and\ \bibinfo {author} {\bibfnamefont {D.}~\bibnamefont
  {Errandonea}},\ }\href {https://doi.org/10.1103/PhysRevB.86.104112}
  {\bibfield  {journal} {\bibinfo  {journal} {Phys. Rev. B}\ }\textbf {\bibinfo
  {volume} {86}},\ \bibinfo {pages} {104112} (\bibinfo {year}
  {2012})}\BibitemShut {NoStop}%
\bibitem [{\citenamefont {Pisani}\ \emph {et~al.}(2007)\citenamefont {Pisani},
  \citenamefont {Valent\'{\i}}, \citenamefont {Montanari},\ and\ \citenamefont
  {Harrison}}]{jrn:pisani07}%
  \BibitemOpen
  \bibfield  {author} {\bibinfo {author} {\bibfnamefont {L.}~\bibnamefont
  {Pisani}}, \bibinfo {author} {\bibfnamefont {R.}~\bibnamefont
  {Valent\'{\i}}}, \bibinfo {author} {\bibfnamefont {B.}~\bibnamefont
  {Montanari}},\ and\ \bibinfo {author} {\bibfnamefont {N.~M.}\ \bibnamefont
  {Harrison}},\ }\href {https://doi.org/10.1103/PhysRevB.76.235126} {\bibfield
  {journal} {\bibinfo  {journal} {Phys. Rev. B}\ }\textbf {\bibinfo {volume}
  {76}},\ \bibinfo {pages} {235126} (\bibinfo {year} {2007})}\BibitemShut
  {NoStop}%
\bibitem [{\citenamefont {Tomczak}(2007)}]{phd:tomczak}%
  \BibitemOpen
  \bibfield  {author} {\bibinfo {author} {\bibfnamefont {J.~M.}\ \bibnamefont
  {Tomczak}},\ }\emph {\bibinfo {title} {{Spectral and Optical Properties of
  Correlated Materials.}}},\ \href
  {https://pastel.archives-ouvertes.fr/pastel-00003163} {\bibinfo {type}
  {Theses}},\ \bibinfo  {school} {{Ecole Polytechnique X}} (\bibinfo {year}
  {2007})\BibitemShut {NoStop}%
\bibitem [{\citenamefont {Perdew}\ \emph {et~al.}(1995)\citenamefont {Perdew},
  \citenamefont {Savin},\ and\ \citenamefont {Burke}}]{jrn:perdew95}%
  \BibitemOpen
  \bibfield  {author} {\bibinfo {author} {\bibfnamefont {J.~P.}\ \bibnamefont
  {Perdew}}, \bibinfo {author} {\bibfnamefont {A.}~\bibnamefont {Savin}},\ and\
  \bibinfo {author} {\bibfnamefont {K.}~\bibnamefont {Burke}},\ }\href
  {https://doi.org/10.1103/PhysRevA.51.4531} {\bibfield  {journal} {\bibinfo
  {journal} {Phys. Rev. A}\ }\textbf {\bibinfo {volume} {51}},\ \bibinfo
  {pages} {4531} (\bibinfo {year} {1995})}\BibitemShut {NoStop}%
\end{thebibliography}%
\end{document}